\documentclass[sigconf]{acmart}

\usepackage[utf8]{inputenc}
\usepackage[T1]{fontenc}
\usepackage{textgreek}
\usepackage{multirow}
\usepackage{subcaption}
\usepackage{xcolor,colortbl}
\definecolor{gray}{rgb}{0.1,0.1,0.1}
\usepackage{graphicx}
\usepackage{tabularx}
\usepackage{longtable}
\usepackage{enumitem}
\usepackage{amsmath}
\usepackage{array}
\usepackage{booktabs}
\usepackage{makecell}
\usepackage{framed}
\usepackage{ragged2e}

\AtBeginDocument{%
  \providecommand\BibTeX{{%
    \normalfont B\kern-0.5em{\scshape i\kern-0.25em b}\kern-0.8em\TeX}}}

\copyrightyear{2026}
\acmYear{2026}
\setcopyright{cc}
\setcctype{by}
\acmConference[DIS Companion '26]{Designing Interactive Systems Conference}{June 13--17, 2026}{Singapore, Singapore}
\acmBooktitle{Designing Interactive Systems Conference (DIS Companion '26), June 13--17, 2026, Singapore, Singapore}
\acmDOI{10.1145/3802974.3807979}
\acmISBN{979-8-4007-2632-3/2026/06}

\begin{document}
\title{Designing Youth Social Media through Problem Space Attunement}

\author{JaeWon Kim}
\affiliation{%
  \institution{The Information School, University of Washington}
  \city{Seattle}
  \country{USA}}
\email{jaewonk@uw.edu}

\begin{abstract}
Social media is central to how young people maintain relationships, develop identity, and access communities, yet dominant platform designs often leave youth feeling disempowered rather than supported. My dissertation argues that youth social media design is shaped by three forms of problem-space misattunement. \textit{Conceptual misattunement} occurs when the language of ``social media'' anchors participants to existing platforms' interaction templates. I address this through a Fictional Inquiry design workshop that frees youth from preconceived notions of social media by having them brainstorm ways to ``magically connect with remote wizard friends'' rather than ideas for ``social media.'' \textit{Definitional misattunement} occurs when researchers define what ``better'' means on youth's behalf. I address this through a Discord-based asynchronous community that supports youth-led collective inquiry. \textit{Evaluative misattunement} occurs when participants are asked to judge static or hypothetical designs. I address this through an ego-anchored, LLM-agent simulation sandbox. Together, these studies develop youth-grounded criteria and design directions for relationally supportive social media.
\end{abstract}




\maketitle

\section{Introduction and Motivation}
Social media can support central aspects of young people's development, including friendship, identity development, and access to information about careers, politics, and activism~\cite{kim2025discord, ellison2007, fiesler2019}. Yet in policy, education, and public discourse, social media is increasingly framed less as infrastructure for connection than as a threat from which young people should be protected. These concerns are understandable: mainstream platforms have converged around similar engagement-driven designs, and these designs can undermine youth well-being~\cite{sundaramnoyear, pitt2021SocialMedia, pardes2020}.

Protective responses, however, can create problems of their own when they treat withdrawal from social media as the primary path to safety. Prior work has shown that the emphasis on harm can generate dysfunctional levels of privacy-related concern in young people: anxiety that degrades quality of life without meaningfully improving safety or leading to constructive action~\cite{kim2025privacy}. Young people themselves often describe blanket restrictions as unrealistic and unhelpful. Because peer relationships are developmentally central, exclusion from social media can also mean exclusion from peer life; youth therefore need opportunities to build resilience and shape healthier norms, not only rules that remove access~\cite{kim2025trust, kim2024Not, weinstein2022}.

The burden of managing these tensions currently falls largely on individual users, and youth are especially underserved~\cite{weinstein2022, kim2025privacy}. Young people often do not feel empowered to make choices that align with their preferences because platform norms, themselves shaped by design, are misaligned with youth needs~\cite{kim2025privacy, kim2025trust}. Involving young people in design research is therefore critical~\cite{wisniewski2015, wisniewski2012, wisniewski2018}. Yet when they participate, they may reproduce the adult-driven narratives they have been taught rather than articulating their own felt needs~\cite{kim2026hogwarts, hiniker2016}. This creates a methodological problem: simply involving youth in the design process does not guarantee that the problem space is framed or solutions aligned on youth's own terms and actual needs.

My dissertation asks how design research can better attune to youth when defining and addressing the problem space of social media design. I use \textit{problem space attunement} to name the work of aligning research questions, design criteria, and evaluation methods with young people's lived experience, rather than with adult assumptions or existing platform templates. Attunement is not a new method I propose; it is a practical orientation that guides how I adapt existing approaches such as Fictional Inquiry~\cite{dindler2007}, participatory design, and LLM-based simulation to this specific problem space. The three studies are sequential. Fictional Inquiry widens the conceptual frame; participatory collective inquiry lets youth define what better social media should be for; and an LLM-based simulation sandbox helps youth evaluate how design choices might play out in familiar social contexts. Together, these studies produce empirical findings about youth's relational needs, youth-articulated design goals, and evaluated design directions for social media that supports young people.

\section{Background}
Two bodies of work ground my approach. The first concerns how problem framing shapes design outcomes. Sch\"{o}n~\cite{schon1983} argued that the most consequential act in design is problem \textit{setting}, meaning deciding what counts as the problem, rather than problem solving. Dorst~\cite{dorst2011} extended this with \textit{frame creation}, showing how new ways of understanding a situation can open previously invisible solution paths. In youth social media research, the problem is often framed around harms such as addiction, comparison, and misinformation. This framing yields useful interventions, but it can also overlook the relational, developmental, and identity needs that motivate youth engagement with social technology in the first place.

The second body of work concerns the difficulty of truly accessing youth experience through conventional methods. Developmental sensitivity to peer evaluation can make young people especially prone to social desirability bias and demand characteristics in research settings~\cite{valkenburg2022}. Youth privacy concerns are primarily relational, including concerns about friends screenshotting content or about reaching unintended audiences, rather than purely informational. Yet research instruments often foreground informational privacy~\cite{kim2025privacy, kim2025trust, marwick2011}. The term ``social media'' itself can also constrain imagination by anchoring participants' thinking to a narrow set of platforms rather than to the broader design space of mediated social connection~\cite{kim2026hogwarts}.

I draw on Stern's~\cite{stern1985} concept of \textit{affect attunement}, in which caregivers perceive and respond to experience that cannot yet be explicitly stated, as a sensitizing analogy for the researcher's task. Stern observed that attuned caregivers respond not by mirroring behavior literally but by matching its underlying contour: the intensity, rhythm, or shape of what the child is feeling. The analogy is apt because youth have rich relational experiences with social media, but the available language for describing them, whether platform interface labels, harm-centered public discourse, or predefined research constructs, is often borrowed and may not fit. Attunement names the methodological work of designing inquiry so that pre-articulate experience can surface before it is filtered through inherited frames.

\section{Three Forms of Misattunement in Youth Social Media Design}
I identify three linked forms of misattunement in youth social media design research and address each through a dedicated study.

\subsection{Conceptual Misattunement: Expanding the Meaning of Social Media through Fictional Inquiry}
When mental models of social media default to a handful of mainstream platforms like Instagram or TikTok, participants tend to envision only incremental fixes that react to the recurring frustrations of those platforms: a cleaner feed, stronger moderation, or fewer metrics that encourage social comparison. This is a form of cognitive and design fixation~\cite{jansson1991} that arises naturally from how mainstream platforms have converged on a narrow set of design patterns, all optimized around similar incentive structures. The framing constrains the problem space before inquiry even begins.

To break this constraint, I conducted co-design workshops using the Fictional Inquiry (FI) method~\cite{dindler2007} with 23 youth and young adults aged 15--24~\cite{kim2026hogwarts}. Participants imagined how Hogwarts students might connect with remote friends using any magical powers they wished. We chose the Harry Potter setting~\cite{harris2023} because, unlike other well-known magical worlds such as Lord of the Rings~\cite{lordoftherings} or Star Wars~\cite{starwars}, its central characters are school students navigating ordinary relational concerns: friendships, conflict, identity, and belonging. This gave participants a rich source of inspiration grounded in everyday social life. Reframing the task as ``connecting remotely via magic'' rather than designing ``social media'' suspended both the vocabulary and the technical constraints of existing platforms, freeing participants from the preconceptions that the loaded term ``social media'' carries and letting them reason from felt relational needs.

Participants' visions diverged markedly from what co-design studies typically produce. When initially asked what they wanted from better social media, they named goals framed against existing platforms: less doomscrolling, fewer hours on reels, less performative self-presentation. Once freed from that framing, however, they shifted toward what genuine connection with friends actually feels like. They imagined coexisting ambiently in a shared room while listening to music, designing personal houses that reflected their ideal selves, and living in neighborhoods structured so that closer friends lived closer by. They imagined connecting with friends in organic, low-intensity, playful ways, such as walking through the neighborhood together. These visions were closer to Minecraft or Discord than to Instagram. Their designs converged on four themes that articulated relational needs current platforms often marginalize: intuitive spatial navigation, authentic self-expression beyond quantified metrics, granular and context-sensitive access, and playful pathways for friendship development. This study showed that the design space changes substantially when youth are not constrained by the vocabulary of existing platforms.

The study also produced a methodological finding. Participants entered the workshop wanting foolproof ways to stop doomscrolling and spend less time on social media, and described social media's future as doomed. They left describing the exercise as hopeful and agentic, in sharp contrast to the ``doom narrative'' that research and policy settings often reinforce~\cite{kim2025positech}.

\subsection{Definitional Misattunement: Clarifying ``Better'' Social Media through Participatory Collective Inquiry}
FI clarifies which constraints on social media's future are real and which are instead artifacts of how the problem space has been defined. By showing that the solution space is not fixed but largely underexplored, it gives youth room to imagine. What it does not address, however, is who defines the criteria for evaluating those imaginings. Researchers typically pursue ``better'' social media by addressing discrete harms that current platforms have surfaced. This work is valuable, but each problem is one facet of a much larger landscape, and treating harms in isolation produces a partial picture. More elaborate privacy controls, for example, will not work if they collide with peer norms, with youth's lack of trust in the platform, or with the interpersonal trust between users in the first place. Resolving discrete harms also does not guarantee that social media becomes supportive or meaningful. The absence of negative outcomes is not the presence of good, particularly for the identity and relationship work that is central to young people's needs~\cite{kim2024positech, davis2025, landesman2024}.

Platform designers, meanwhile, optimize for engagement metrics that prior work has shown to be misaligned with what youth actually want~\cite{davis2025, landesman2024}. In both research and commercial design, the terms of the problem are set by those metrics rather than by youth themselves: young people are invited to generate solutions but rarely to define what the design is for. This is definitional misattunement.

To address it, I am conducting a participatory collective inquiry using an Asynchronous Remote Community (ARC) approach~\cite{maclean2015} as the first phase of a broader co-design study. Over multiple weeks, youth participants engage in semi-structured discussions around questions about goals and criteria: What would ``better'' social media actually feel like? What metrics or felt experiences would confirm that a design is working? What edge cases appear beneficial on the surface but are not? Which apparently harmful practices serve a genuine need? Which design goals matter most? A bot scaffolds discussions by posting prompts and summarizing contributions, so youth can drive the inquiry with minimal adult researcher intervention.

The ARC format is deliberate. Asynchronous discussion over days reduces the social pressure that can amplify demand characteristics in real-time group settings. It also allows participants to build on each other's reasoning iteratively, producing collective definitions rather than individual reactions. Finally, because the ARC itself is an instance of online community-building, participants refine their thinking through firsthand experience of the very dynamics they are studying, developing a sharper sense of how platform design choices shape interaction and a stronger literacy for evaluating social media design.

This phase will produce a youth-articulated framework of goals, criteria, and boundary conditions for relationally supportive social media. That framework will then guide both the design of the simulation system and the evaluation of design outcomes.

\subsection{Evaluative Misattunement: Grounding Design Judgment through a Simulation-Based Sandbox}
Even with the right conceptual frame and youth-defined criteria, a third barrier remains. Social media outcomes are emergent: whether a platform feels safe, authentic, or supportive depends on how features interact with user behavior, norm formation, and social context. These dynamics often reveal themselves only over time and at scale, and shift with conditions such as the composition of the initial user base. Prior work shows that survey ratings of proposed features exhibit ceiling effects with limited variance~\cite{kim2025privacy}, and co-design interviews tend to surface optimistic scenarios while obscuring risks, particularly when participants or the adult researchers they speak with are generating those scenarios on the spot. These are not failures of youth judgment; the task of anticipating collective social dynamics from a static description is inherently difficult for anyone.

To scaffold this reasoning, I am building a simulation sandbox that populates a social media environment with LLM-powered agents grounded in each participant's actual social network. Participants describe people they interact with, such as close friends, acquaintances, and family, along dimensions identified as meaningful through the collective inquiry phase. The system then generates agents calibrated to these descriptions. These agents post, react, share, ignore, or violate boundaries under different platform configurations.

The system is \textit{ego-anchored}: agents represent people the participant actually knows, enabling credibility judgments grounded in lived experience rather than hypothetical reasoning. Crucially, the simulation exposes not only what agents do but also their inner monologues, including the agent that mirrors the participant themselves, giving participants a concrete basis for judgment. The goal is not to predict real people's behavior or to replace human evaluation. Instead, the simulation functions as a reasoning scaffold, surfacing otherwise invisible phenomena: the decision not to post after weighing perceived norms, the silent judgment of a friend's content, the perceived judgment from others that shapes how one self-presents.

This process does more than involve youth in evaluation. By letting them observe how specific design choices ripple through familiar social dynamics, it builds a form of design literacy: a sharper sense of how platform configurations shape the everyday decisions they make online, and a stronger footing from which to articulate what they want changed. Evaluation also becomes an inquiry into nuanced, complex, realistic, evolving social dynamics rather than a reaction to a flat, static artifact.

\section{Expected Contributions}
My dissertation will contribute four outcomes. First, it provides empirical findings on what youth seek from social media when freed from the conceptual constraints of existing platforms. Second, it develops a youth-articulated framework of design goals, evaluation criteria, and boundary conditions for relationally supportive social media. Third, it evaluates a simulation tool for co-design that makes design evaluation richer and more nuanced. Fourth, it synthesizes concrete design directions for social media grounded in youth perspectives across all three studies. Fifth, through a field deployment of a platform built on these directions, it empirically tests whether attunement-grounded research produces designs that serve youth better in practice.

Initial deployment results are promising: participants posted roughly twice as often as on a control version of the platform, even while receiving fewer reactions, and reported feeling more intrinsically motivated to share and more curious about others' contributions. The larger contribution is less a new standalone method than an orientation: a way of selecting, adapting, and augmenting existing methods so that the problem space and the solutions developed within it remain aligned with the people being designed for, especially when researchers and participants inhabit different felt experiences, as adults and youth often do.

\section{Planned Next Steps}
I have passed my general examination and completed my dissertation proposal. The Fictional Inquiry study is complete, and the resulting manuscript is currently in preparation. I am now building the simulation platform and developing a course, \textit{Designing Social Media Futures}, that will help form the youth research community needed for the ARC and simulation studies. The course will bring together youth participants across Korea and the United States, creating a sustained setting for cross-cultural discussion, collective inquiry, and design evaluation.

Data collection for the collective inquiry phase is planned for late 2026, during the course. The ARC study and simulation study will be conducted in parallel: the ARC discussions will help participants articulate goals, criteria, and boundary conditions for better social media, while the simulation activities will let them evaluate how those criteria play out in dynamic, socially grounded scenarios. A field deployment of a platform built on the resulting design directions will then test whether attunement-grounded research produces designs that serve youth better in practice. After data collection, I will synthesize findings across all three studies to develop final design implications for relationally supportive youth social media. I expect to defend the dissertation in 2027.

\begin{acks}
I would like to acknowledge the CERES Network, University of Washington Global Innovation Funds (GIF), and Student Technology Funds (STF), which provided support for this work. This work was also funded in part by the Paul G. Allen School of Computer Science \& Engineering Endowed Fund for Excellence and a gift from Google. I thank my advisor, Alexis Hiniker, and committee members Katie Davis, Casey Fiesler, Amy X. Zhang, and Sean Munson for their guidance throughout this work.
\end{acks}

\bibliographystyle{ACM-Reference-Format}
\bibliography{references, newrefs, references1, references2, references3}

\end{document}